\documentclass[draftcls,onecolumn,12pt]{IEEEtran}
\usepackage{amsmath}
\usepackage{amssymb}
\usepackage{graphicx}
\usepackage{url}
\usepackage{multirow}
\usepackage{epstopdf}
\usepackage{fmtcount}
\usepackage{algorithmic}
\usepackage{algorithm}
\usepackage[tight]{subfigure}
\graphicspath{{figures/}}
\begin{document}
\title{Flip-OFDM for Unipolar Communication Systems}

\author{Nirmal Fernando,
        Yi~Hong,
        Emanuele~Viterbo
\thanks{\scriptsize N. Fernando, Y. Hong and E. Viterbo are with
the Department of Electrical and Computer System Engineering, Monash University, Clayton, VIC
3800, Australia, E-mail: $ \tt Nirmal.Fernando, \tt Yi.Hong, \tt Emanuele.Viterbo  $@$\tt
monash.edu$. }}

\date{\today}

\maketitle
\begin{abstract}
Unipolar communications systems can transmit information using only real and positive signals.
This includes a variety of physical channels ranging from optical (fiber or free-space),
to RF wireless using amplitude modulation with non-coherent reception, to baseband single wire
communications.
Unipolar OFDM techniques enable to efficiently compensate frequency selective distortion in the
unipolar communication systems.
One of the leading examples of unipolar OFDM is asymmetric clipped optical OFDM (ACO-OFDM)
originally proposed for optical communications.
Flip-OFDM is an alternative approach that was proposed in a patent,
but its performance and full potentials have never been investigated in the literature.
In this paper, we first compare Flip-OFDM and ACO-OFDM, and show that both techniques have
the same performance but different complexities (Flip-OFDM offers 50\% saving).
We then propose a new detection scheme, which enables to reduce the noise at the Flip-OFDM
receiver by almost 3dB.
The analytical performance of the noise filtering schemes is supported by the simulation results.
\end{abstract}


\section{Introduction}
Unipolar communications systems can transmit information using only {\em real} and {\em positive} signals.
Common examples of unipolar communication systems include optical communications (fiber and free space)
\cite{kahn_wireless_1997,armstrong_ofdm_2009},
non-coherent wireless communications \cite{Goldsmith05},
and baseband digital communications over a single wire \cite{john_digital_communications}.
Channel dispersion or multipath fading may cause the intersymbol interference and
degrade the performance of such unipolar communication systems.
To compensate these effects, unipolar orthogonal frequency division multiplexing (OFDM),
can be used.
Three different unipolar OFDM techniques are described below.
\begin{itemize}
\item
DC-offset OFDM (DCO-OFDM) \cite{armstrong_ofdm_2009}, known as the traditional unipolar OFDM
technique, uses Hermitian symmetry property with a DC-bias to generate a {\em real} and {\em positive}
time domain signal. However, the DC bias depends on the peak-to-average-power ratio (PAPR) of
the OFDM symbol. Since OFDM has a high PAPR, the amplitude of the DC bias is generally
significant. It was shown in \cite{armstrong_comparison_2008} that the requirement of large DC-bias
makes DCO-OFDM optically power inefficient. The use of lower DC bias can
lead to frequent clipping of negative parts of the time-domain signal.
This can cause inter-carrier interference and create out-of-band optical power.
\item
Asymmetrically clipped optical OFDM (ACO-OFDM) was
proposed in \cite{armstrong_power_2006} and does not require any DC bias.
ACO-OFDM only uses odd subcarriers to transmit information symbols,
and the negative part of the time-domain signal is clipped.
It was shown in \cite{armstrong_power_2006} that this clipping does not distort information symbols
in odd subcarriers, although their amplitudes are scaled by half.
In \cite{armstrong_comparison_2008,armstrong_performance_2006, xia_li_channel_2007},
the performance of ACO-OFDM was compared to other modulation schemes such as on-off keying
and DC-biased OFDM (DC-OFDM); and it was shown that ACO-OFDM has better power efficiency
over optical wireless channels \cite{armstrong_comparison_2008}. Performance of ACO-OFDM can be
further improved by using bit loading and diversity
combining schemes, as discussed in \cite{wilson_digital_2008, wilson_transmitter_2009, liang_chen_diversity_2009}.
Different from the above comparison over optical wireless channels, in \cite{Khan_2011},
the power efficiency comparison between ACO-OFDM, on-off keying, and DC-OFDM
are presented specifically for single-mode fiber optical communications.
\item
An alternative unipolar OFDM technique to ACO-OFDM
was proposed in \cite{yong_modulation_2007} and has
been widely ignored in the open literature to the best of our knowledge. We
name this technique as Flip-OFDM.
In Flip-OFDM, positive and negative parts are extracted from the real bipolar OFDM
symbol generated by preserving the Hermitian symmetry property of transmitted
information symbols. Then the polarity of negative parts are inverted before
transmission of both positive and negative parts in two consecutive OFDM symbols.
Since the transmitted signal is always positive, Flip-OFDM is indeed an unipolar
OFDM technique that can be used for unipolar communications.
\end{itemize}

In this paper we provide three main contributions:
\begin{itemize}
  \item we review and analyze Flip-OFDM in
a general setting of unipolar communication systems;
  \item we modify original Flip-OFDM and compare key system parameters with ACO-OFDM including
  including spectral efficiencies, bit error rate (BER) performance and complexity;
  \item we propose a new detection scheme for Flip-OFDM and analyze
its BER performance. We show, both analytically and by simulations, that a significant BER improvement can be
achieved using the proposed detection scheme in Flip-OFDM.
\end{itemize}

The rest of the paper is organized as follows. In Section II, we define the unipolar
communication systems that can be benefit from unipolar OFDM techniques.
In Section III, we introduce Flip-OFDM and ACO-OFDM and compare their key system
parameters including spectral efficiencies, BER performance and hardware complexities.
In Section IV, we propose a new detection scheme for Flip-OFDM. We also analyze
the performance of Flip-OFDM using the new detection scheme.
Finally, conclusions are drawn in Section V.

\section{Unipolar communication model}

\begin{figure}[t!]
        \centering%
        \includegraphics[scale=.60]{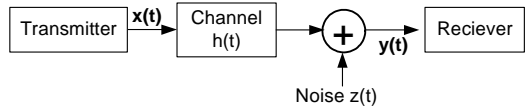}
        \caption{Equivalent model for unipolar communication system.}
        \label{IR_BB_channel_model}
\end{figure}

A non-coherent communication system can be modeled as a linear baseband equivalent system,
as shown in Fig. \ref{IR_BB_channel_model}. Let $x(t)$, $h(t)$ and $z(t)$ represent the transmit
signal (e.g. intensity or amplitude signal), the channel impulse response,
and the noise component, respectively.
Then the non-coherent communication is said to be unipolar if the following two conditions are satisfied:
\begin{enumerate}
\item
$x(t)$ is real and $x(t)\geq 0$ for all $t$.
\item
if the equivalent received signal $y(t)$ can be modeled as
     \begin{equation}
                \label{base_band_channel_model}
                       y(t) = h(t) \otimes x(t) + z(t)
      \end{equation}
where $\otimes$ represents convolution, $h(t) \geq 0$ for all $t$ and $z(t)$ is Gaussian noise with zero mean
and power $\sigma^2_z$.
\end{enumerate}
If the channel is normalized such that $\int^{+\infty}_{-\infty} |h(t)|^2\, dt= 1$, then
the \emph{equivalent signal-to-noise ratio} ($\text{SNR}$) is defined as
    \begin{eqnarray}
        \label{elec_snr_def}
        \text{SNR} = \frac{E[x^2(t)]}{\sigma^2_z}
    \end{eqnarray}
where $E[\cdot]$ is the expectation operator.

Note that such an equivalent baseband model can represent the process of modulation and demodulation of
bandpass signals transmitted over the physical channels.
Following are common examples for such unipolar communication systems:
\begin{itemize}
\item
\emph{Optical communications (fiber or free space)} --
The unipolar information carrying signal $x(t)$ modulates the optical intensity of a LED or a laser.
Since the intensity can not be negative and the photodetector can not recover the phase of the optical
carrier at the receiver, the equivalent optical channel can be modeled with a unipolar $h(t)$
\cite{kahn_wireless_1997, carruthers_modeling_1997, jungnickel_physical_2002}.
\item
\emph{Non-coherent RF wireless} --
In a typical RF communication, the information is transferred through amplitude and phase of a carrier signal.
At the receiver, complex baseband processing is necessary to recover the phase information.
In non-coherent RF wireless, only amplitude modulation with a unipolar modulating signal can be used.
This enables to apply a simple envelope detection so that
the equivalent channel can be modeled with a unipolar $h(t)$.
\item
\emph{Baseband digital communication} --
A baseband system transmitting data over a single wire (e.g. TTL logic \cite{john_digital_communications})
can only use positive signals (e.g. unipolar NRZ \cite{john_digital_communications}).
The channel can be modeled with an unipolar $h(t)$.
\end{itemize}

\section{Unipolar OFDM Techniques}

In this section, we compare Flip-OFDM and ACO-OFDM in the general setting of unipolar communication systems.


\subsection{Flip-OFDM}

\begin{figure*}[!t]
        \centerline{
        \subfigure[Flip-OFDM transmitter]{\includegraphics[scale=.43]{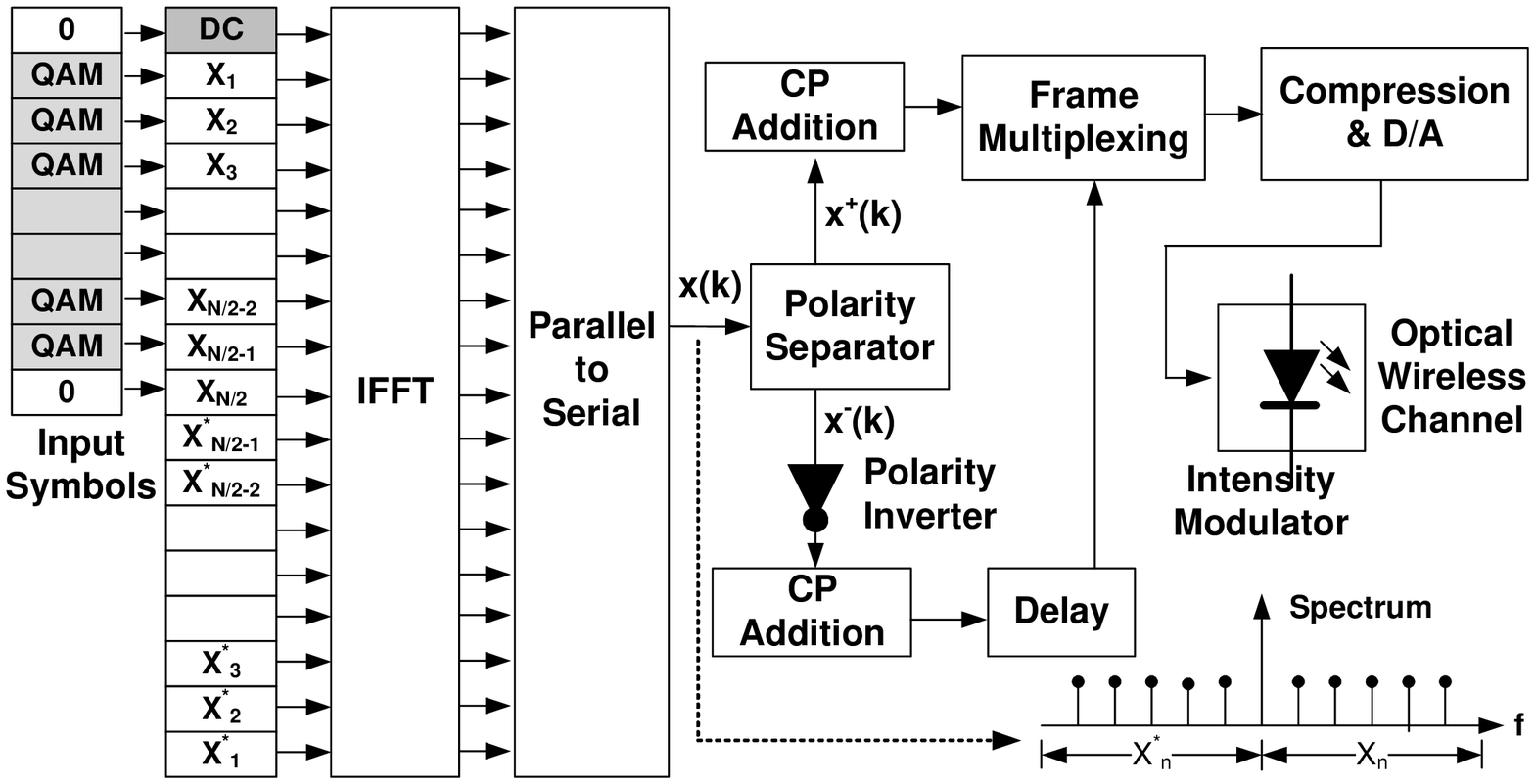}
        \label{flipped_tx}}
        \hfil
        \subfigure[Flip-OFDM receiver]{\includegraphics[scale=.43]{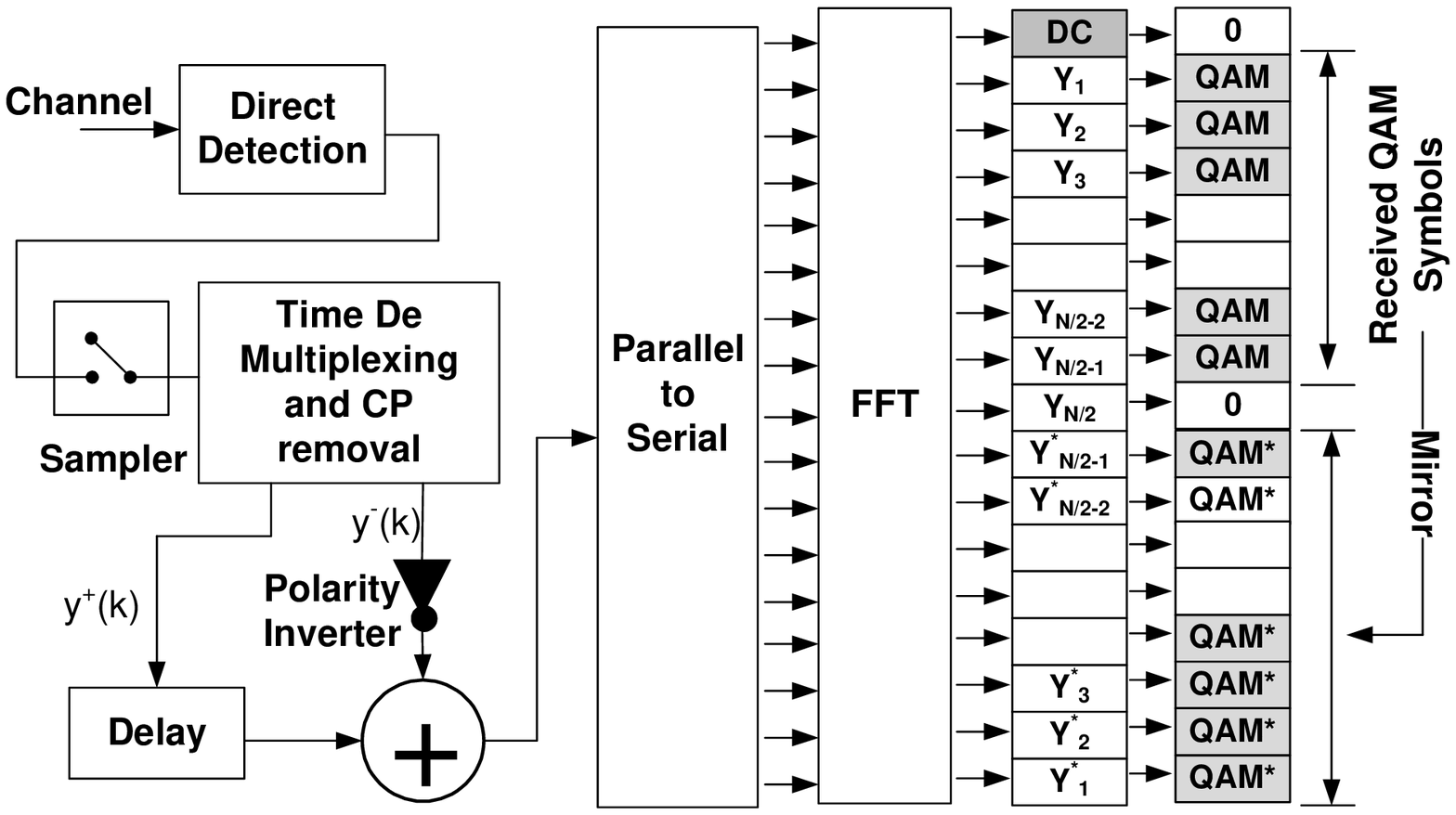}
        \label{flipped_rx}}
        }
        \centerline{
        \subfigure[ACO-OFDM transmitter]{\includegraphics[scale=.45]{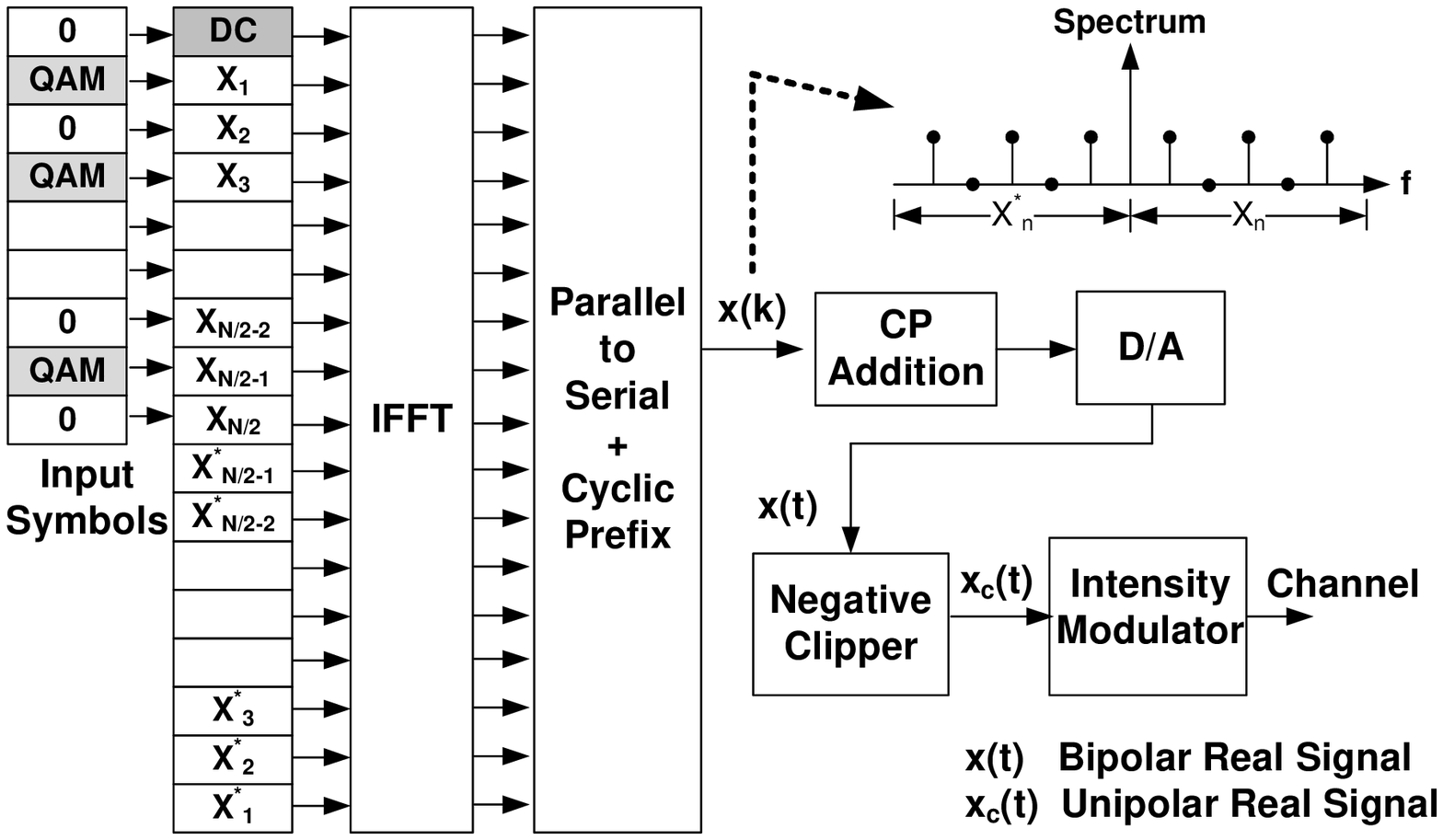}
        \label{aco_tx}}
        \hfil
        \subfigure[ACO-OFDM receiver]{\includegraphics[scale=.45]{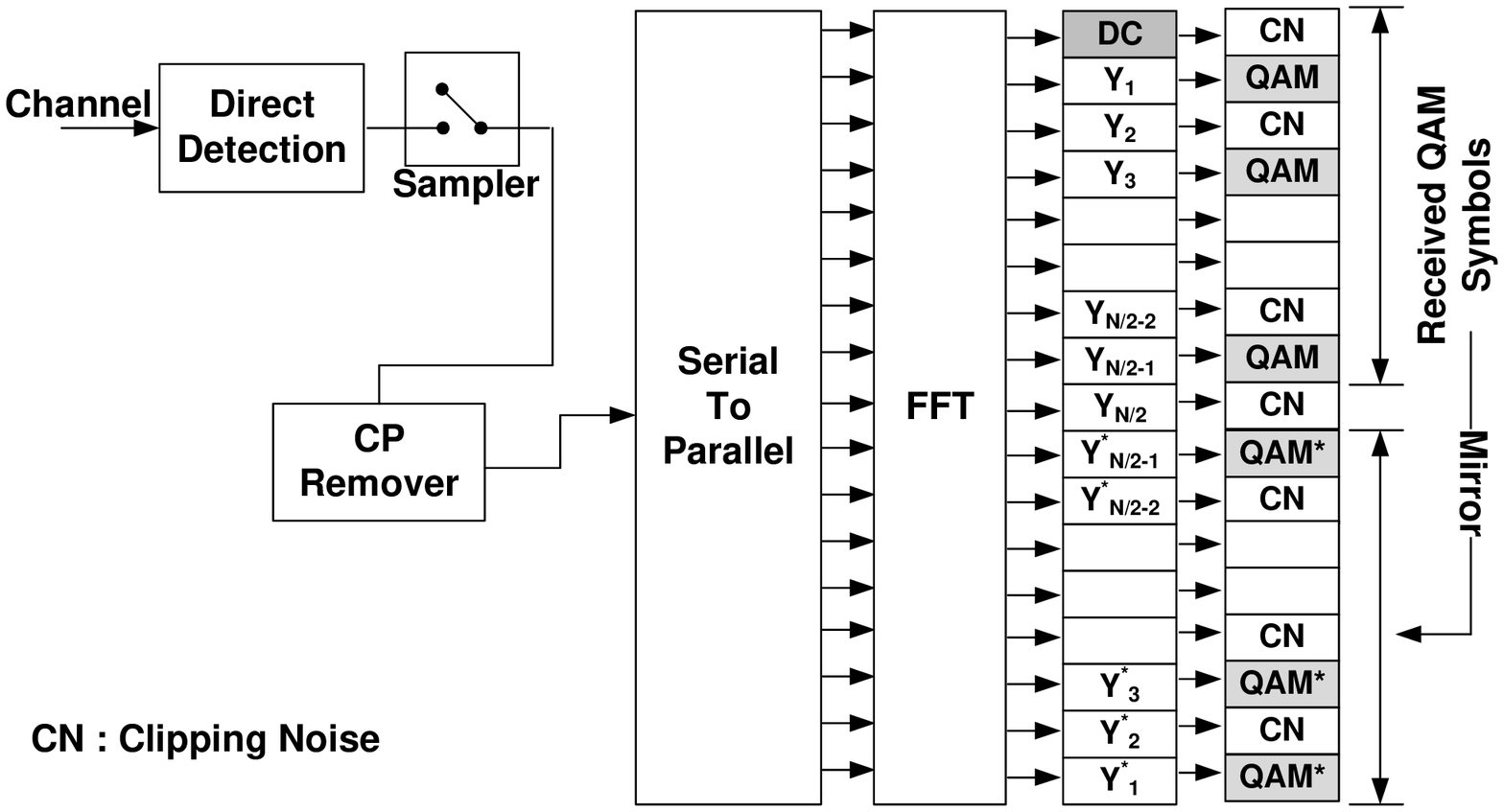}
        \label{aco_rx}}
        }
        \caption{Block Diagrams of Flip-OFDM and ACO-OFDM transmitters and receivers}
        \label{fig_sim}
\end{figure*}

A block diagram of Flip-OFDM transmitter is shown in Fig. \ref{flipped_tx}. Let $X_n$ be the
transmitted QAM symbol in the $n$-th OFDM subcarrier. The output of Inverse Fast Fourier
Transform (IFFT) operation at the $k$-th time instant is given by
    \begin{equation}
        \label{genearl_ofdm_equation}
          x(k)  = \sum_{n=0}^{N-1}{X_n\exp{\left(\frac{j 2 \pi n k}{N}\right)}}
    \end{equation}
where $N$ is the IFFT size and $j^2 =-1$. If the symbol $X_n$ transmitted over each OFDM subcarrier
is independent, the time-domain signal $x(k)$ produced by the IFFT operation is complex.
A real signal can be then obtained by imposing the Hermitian symmetry property
    \begin{equation}
        \label{hermitian_symmetry_lbl}
           X_{n} = X^\ast_{N-n}, \hspace{4mm} n = 0,1,2 ... , N/2-1
    \end{equation}
where $^{\ast}$ denotes complex conjugation. This property implies that half of the OFDM
subcarriers are sacrificed to generate the real time-domain signal. The output of IFFT
operation in (\ref{genearl_ofdm_equation}) can be rewritten as
     \begin{eqnarray}
       \label{HS_ofdm_equation}
       x(k)  & = & X_0 + \sum_{n=1}^{N/2-1}{X_n\exp{\left(\frac{j 2 \pi n k}{N}\right)}} + X_{N/2} \exp{\left(j \pi k\right)} \nonumber  \\
             & + & \sum_{n=N/2+1}^{N-1}{{X^*_{N-n}}\exp{\left(\frac{j 2 \pi n k}{N}\right)}}
     \end{eqnarray}
where $X_0$ is the DC component. To avoid any DC shift or any residual complex component
in the time domain signal, we let
\[
X_{0} = X_{N/2} = 0.
\]
In such a way, the output of the IFFT operation is a real bipolar signal.
We can then decompose the bipolar signal as
\[
x(k)=x^+(k) + x^-(k)
\]
where the positive and negative parts are defined as
    \begin{eqnarray}
     \label{flipping_process}
     \nonumber x^+(k) &=& \begin{cases}
             x(k) & \text{if } x(k) \geq 0 \\
             0 &  \text{otherwise}\\
     \end{cases} \\
     x^-(k) &=& \begin{cases}
             x(k) & \text{if }x(k) < 0 \\
             0 &  \text{otherwise}\\
     \end{cases}
     \end{eqnarray}
and $k = 1,2,... ,N$. These two components are separately transmitted
over two successive OFDM symbols.
The positive signal $x^+(k)$ is transmitted in the first subframe
(positive subframe), while the flipped  (inverted polarity) signal $-x^-(k)$
is transmitted in the second subframe (negative subframe).
Since the transmission is over a frequency selective channel,
the cyclic prefixes composed of $\Delta$ samples are added to
each of the OFDM subframes. Hence, the negative OFDM subframe is delayed
by $(N+\Delta)$ and transmitted after the positive subframe.

The reconstruction of the bipolar OFDM frame and the detection process at the
receiver are illustrated in Fig. \ref{flipped_rx}.
The cyclic prefixes associated with each OFDM subframe are removed.
Then the original bipolar signal is reconstructed as
     \begin{equation}
        \label{regen_bipolar_frm}
        y(k) = y_1(k) - y_2(k)
     \end{equation}
where $y_1(k)$ and $y_2(k)$ represent the time-domain samples received in the positive and
negative subframes, respectively. The Fast Fourier Transform (FFT) operation is
performed on the bipolar signal to detect the transmitted QAM information symbols.

\subsection{ACO-OFDM}
A block diagram of an ACO-OFDM transmitter is shown in Fig. \ref{aco_tx}.
At the transmitter, the QAM information symbols are first mapped onto
the first half of the odd subcarriers, $X_{2n+1}$,  where $n = 0,1,2,... ,N/4-1$.
The even subcarriers are set to zero, i.e.
     \begin{equation}
        \label{ACO_condition}
        X_{2n} = 0, ~~~n = 0,1,2,... ,N/2
     \end{equation}

Using the above equation, the DC component and the symbol of the $\frac{N}{2}$-th subcarrier become zero.
The Hermitian symmetry property in (\ref{hermitian_symmetry_lbl}) is used to construct a real signal.
After the IFFT operation, the time-domain OFDM symbol $x(k)$ can be computed using (\ref{HS_ofdm_equation})
and has an odd symmetry property
     \begin{equation}
        \label{symmetic_property}
        x(k) = - x\left(k+ \frac{N}{2}\right).
     \end{equation}
This allows clipping of the negative time samples of $x(k)$ right after the DA conversion at
the transmitter without destroying the original information.
The clipped signal $x_c(t)$ is a unipolar signal, defined as
      \begin{equation}
      \label{time_domain_clipping}
       x_c(t) =  \begin{cases}
             x(t) & \text{if } x(t)\geq 0\\
             0 & \text{Otherwise}.\\
             \end{cases}
      \end{equation}
The cyclic prefix is then added to the clipped unipolar OFDM symbol before the transmission.

The direct detection (DD) of the received signal $y(t)$ is performed at the receiver,
as illustrated in Fig. \ref{aco_rx}. The cyclic prefix of the OFDM symbol is
removed and the serial-to-parallel conversion is performed. The FFT operation is
performed and finally the QAM information symbols contained in the odd subcarriers
can be detected.

\subsection{System Comparisons}
In this subsection, we compare the key system parameters of Flip-OFDM and ACO-OFDM.
For a fair comparison, we use the same channel model with the same delay spread.

\begin{figure}[t!]
        \centering %
        \includegraphics[scale=.45]{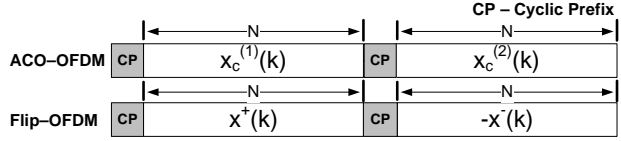}
        \caption{ OFDM symbol structure used to compare Flip-OFDM and ACO-OFDM.
        We assume FFT and IFFT sizes of both ACO-OFDM and Flip-OFDM are the same.
        $N$ denotes the FFT and IFFT size for each case.}
        \label{ofdm_frame_struct}
\end{figure}

{\em Modification to Flip-OFDM}: Note that the original Flip-OFDM \cite{yong_modulation_2007}
uses the compression of time samples to be compliant with the standard bipolar OFDM symbol
length. Given the same bipolar OFDM symbol length for both systems,
this compression in Flip-OFDM leads to half length of each cyclic prefix,
when compared to that of ACO-OFDM.
This implies that both systems have different capabilities to combat delay
spread distortion of the channel.
Here, we do not compress the time scale and two consecutive OFDM symbols of Flip-OFDM
have the same bandwidth and the same cyclic prefix as those of ACO-OFDM, as shown in
Fig. \ref{ofdm_frame_struct}.

{\em Spectral Efficiency}: In ACO-OFDM, each OFDM symbols (i.e. $x_c^{(1)}$ and $x_c^{(2)}$)
has $N/4$ information symbols. However, in Flip-OFDM, even though each symbol has twice of
the number of information symbols (i.e. $N/2$), both positive and negative OFDM subframes
are required to extract the original transmitted information symbols.
Therefore, the data rates in both schemes are approximately the same.
Given the same bandwidth, the spectral efficiencies of both schemes
are indeed the same.

\begin{figure}[t]
\centering
\subfigure[Flip-OFDM: effects on symbol energy during the flipping and the recombination process]{
\includegraphics[scale=.6]{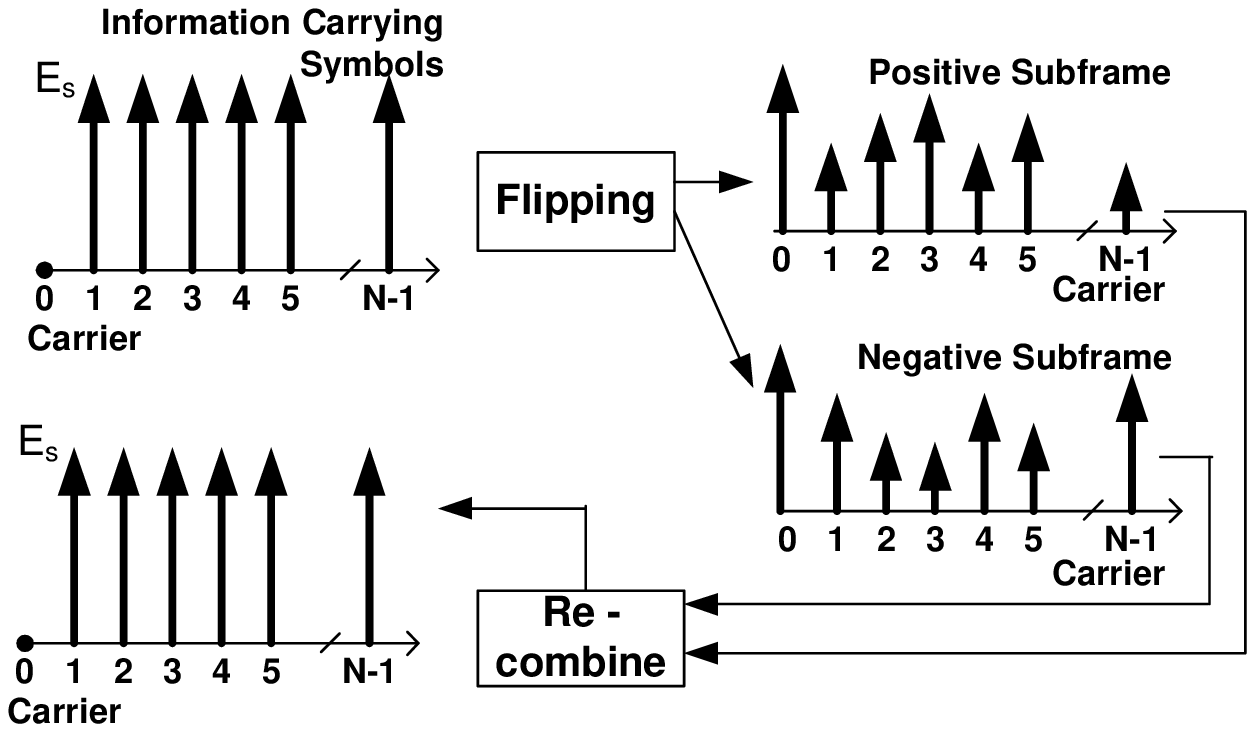}
\label{flip_ofdm_sym_energy}
}
\subfigure[ACO-OFDM: effects on symbol energy due to the asymmetric clipping]{
\includegraphics[scale=.6]{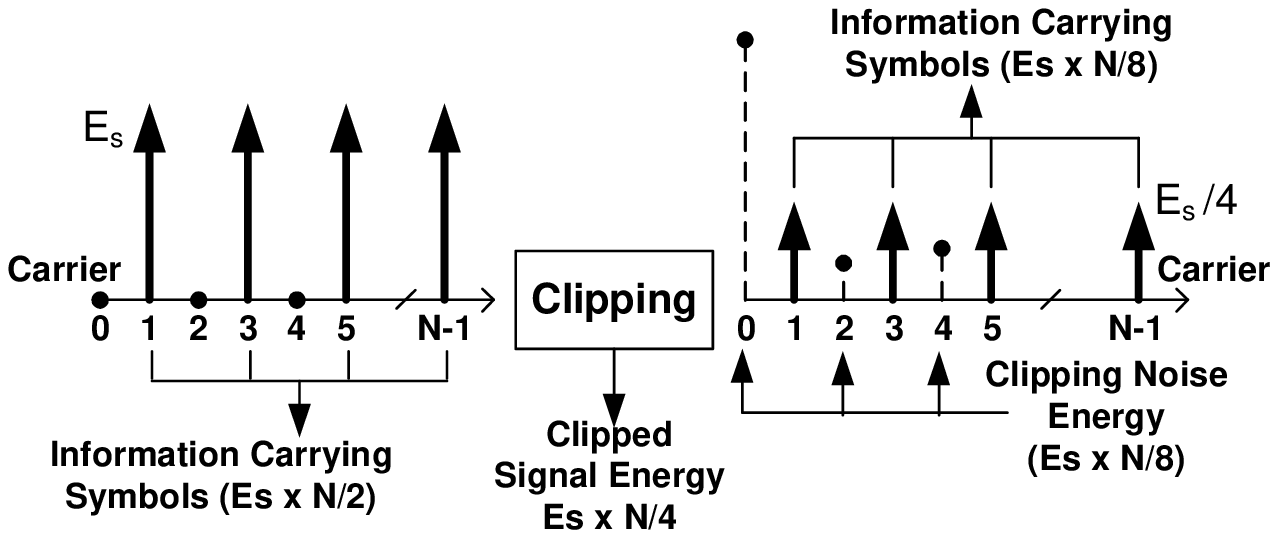}
\label{ACO_ofdm_sym_energy}
}
\label{fig:subfigureExample}
\caption[]{OFDM symbol structure and a comparison of ACO-OFDM and Flip-OFDM symbol energy}
\end{figure}

{\em Symbol Energy}: In Flip-OFDM, the energy of an information symbol
is spread across the positive and negative OFDM subframes during
the flipping process, as shown in Fig. \ref{flip_ofdm_sym_energy}.
However, this spread energy is fully recovered at the receiver by
the recombination of the subframes.

In ACO-OFDM, since the OFDM symbol is symmetric around time axis,
the clipping preserves half of the original signal energy and
scales the amplitude of the original symbols by half
\begin{equation}
          \label{cliping_affect_on_infor_symb}
          X^c_{2n+1} = \frac{1}{2} X_{2n+1}
\end{equation}
where $X^c_{2n+1}$ denotes the information carrying symbol after the
asymmetric clipping process.
Hence, the energy of information carrying symbol is reduced by a
fraction of four, while the clipping has
shifted the other quarter of the signal energy (half of the
signal energy is lost during the clipping process)
into the odd subcarriers, as illustrated in
Fig. \ref{ACO_ofdm_sym_energy}.
This energy in the odd subcarriers is known as
clipping noise \cite{armstrong_power_2006, armstrong_performance_2006}.
Therefore, the energy of an information symbol in Flip-OFDM is
twice the amount of ACO-OFDM for a given transmitted power.

{\em Noise Power}: In Flip-OFDM, the noise power of the Flip-OFDM is
doubled during the recombination of the positive and negative OFDM subframes.
Let $H^+_n$ and $H^-_n$ be the channel responses of $n$-th OFDM subcarrier
over two subframes respectively,
the outputs of the $n$-th OFDM subcarrier in the two subframes are
      \begin{eqnarray}
        \label{flip_rx_positive}
                  Y^+_n  &=& H^+_n X^+_n + Z_n^+\\
        \label{flip_rx_negative}
                  Y^-_n  &=& -H^-_n X^-_n + Z_n^-
      \end{eqnarray}
where $Z_n^+$ and $Z_n^-$ represent the noise components of $n$-th OFDM subcarrier.
Under slow fading characteristics, we can assume the channel is constant over
two consecutive OFDM symbols (i.e. $H^+_n = H^-_n \triangleq H_n$). Then the
addition of (\ref{flip_rx_positive}) and (\ref{flip_rx_negative}) yields
     \begin{eqnarray}
        \label{noise_double_eqn_const_channel}
             R_n  &=& H_n X_n + \{ Z_n^+ + Z_n^-\}.
      \end{eqnarray}
where $R_n$ is the received information symbol. As $Z_n^+$ and $Z_n^-$ are
Gaussian (i.e. $\sim {\cal N}(0,\sigma^2_z)$),
the noise power of received information symbols is $2 \sigma^2_z$.

In ACO-OFDM, since there is no recombination, the received information symbol is given by
      \begin{eqnarray}
        \label{aco_recv_infor_symbol}
             R_{2n+1}  &=& \frac{1}{2} H_{2n+1} X_{2n+1} +  Z_{2n+1}.
      \end{eqnarray}
and the noise power is $\sigma^2_z$, which is half of the amount in Flip-OFDM.

{\em Equivalent SNR}: Since half of the transmitted signal energy is preserved
in ACO-OFDM and the other half is the clipping noise,
the SNRs of both ACO-OFDM and Flip-OFDM are indeed the same.
Using (\ref{elec_snr_def}), the equivalent SNR per received sample is given by
\begin{eqnarray}
        \label{exten_elec_snr_def}
        \text{SNR} = \frac{\sigma^2_x}{2 \sigma^2_z}
\end{eqnarray}
where $\sigma^2_x$ is the transmitted signal power.

{\em Bit Error Rates:} The analytical BER expression for both Flip-OFDM and ACO-OFDM in AWGN channels
can be computed as \cite{john_digital_communications}
\begin{eqnarray}
\label{theroitical_BER_eqn}
P_b  \simeq \frac{2}{\log_2M} \left(1 - \frac{1}{\sqrt{M}} \right) ~ \mbox{erfc}\left(\sqrt{\frac{3}{2(M-1)} \text{SNR}}\right)
\end{eqnarray}
for a rectangular $\emph{M}$-QAM constellation, where $\mbox{erfc}(\cdot)$ is the complementary error function.
The simulated BER performance of Flip-OFDM and ACO-OFDM for the specified optical wireless
channel having strong LOS signal (Directed, has AWGN characteristics \cite{jungnickel_physical_2002})
and multipath propagation signals (Nondirected or Diffused mode),
were compared in \cite{nirmal_2011}. It was shown in \cite{nirmal_2011} that both systems have the same BER performance,
which can be accurately predicted by (\ref{theroitical_BER_eqn}).

{\em Complexity:~} We define complexity as the number of FFT/IFFT operations at the transmitter or the receiver.
A complexity comparison is given in Table \ref{Hardware_complexities}.
At the transmitter, both schemes have nearly the same complexity for a significant value of $N$,
given the IFFT operation at ACO-OFDM is optimized by zeroing half of subcarriers.
However, at the receiver, Flip-OFDM has a 50\% of complexity savings compared to ACO-OFDM.

    \addtocounter{footnote}{1}
    \begin{table}
    \renewcommand{\arraystretch}{1.4}
    \begin{center}
    \caption{Complexity comparison of Flip- and ACO-OFDMs}
    \begin{minipage}{\linewidth} \center
    \begin{tabular}{l|l|l}
    \hline
                          & ACO-OFDM & Flip-OFDM \\
    \hline \hline

             Transmitter & $2~(\frac{N}{2})\log(\frac{N}{2})$     & $N\log(N)$ \\
             Receiver    & $2N\log(N)$                            & $N\log(N)$ \\
    \hline
    \end{tabular}
    \end{minipage}
    \label{Hardware_complexities}
    \end{center}
    \end{table}
\section{Enhanced detection for Flip-OFDM}
\begin{figure}[t!]
        \centering%
        \includegraphics[scale=.54]{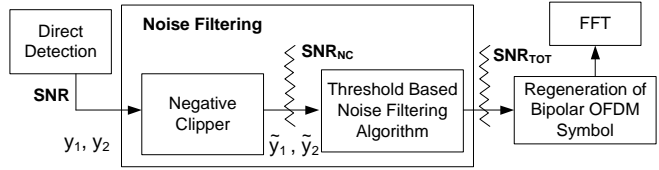}
        \caption{The stages of noise reduction at the Flip-OFDM receiver}
        \label{noise_reduction_stages}
\end{figure}

Assuming the channel has a strong LOS signal (i.e. AWGN),
at the receiver, we introduce a new detection scheme including
two nonlinear noise filtering stages for the time domain samples,
as shown in Fig. \ref{noise_reduction_stages}.
In the first stage of the detection, a negative clipper is placed right after DD.
In the second stage, a threshold based noise filter is used to further improve the BER performance.
Then the preprocessed time samples are sent to the FFT operation for the detection procedure.

\subsection{Negative Clipper}
In unipolar communications, although the transmitted signal is always positive,
the received signal can be negative due to Gaussian noise \cite{kahn_wireless_1997, carruthers_modeling_1997}.
Therefore, negative clipper can be used to force the negative sample to be zero.
This idea was proposed in \cite{wilson_transmitter_2009} for ACO-OFDM;
and can be used in our detection as the first stage of noise filtering.

Recall $y_1(k)$ and $y_2(k)$ in (\ref{regen_bipolar_frm}), i.e.,
the $k$-th received time samples of the positive and negative OFDM subframes.
For simplicity of notation, we omit the index $k$ in the rest of the paper.

Since positive and negative OFDM subframes are defined with same unipolar OFDM frame,
only one sample ($y_1$ or $y_2$) must contain the signal component
$x$ ($x^+$ or $x^-$), i.e. either
\begin{eqnarray}
  \label{noise_samples_in_rcv_signal_pos}
     \nonumber y_1 &=& x^+ + z_1 ~=~ x + z_1\\
               y_2 &=&  z_2
\end{eqnarray}
\noindent{or }
\begin{eqnarray}
  \label{noise_samples_in_rcv_signal_neg}
     \nonumber y_1 &=&  z_1 \\
               y_2 &=&  -x^- + z_2 ~=~ x + z_2
\end{eqnarray}
where $z_1, z_2 \sim {\cal N}(0,\sigma^2_z)$.
For large IFFT sizes, the output produced by the IFFT operation is i.i.d and has a Gaussian distribution
($\sim {\cal N}(0,\sigma^2_x$)). Hence, the pdf of the $x$ is a one sided Gaussian
\begin{eqnarray}
  \label{distribution_of_tx_signal}
   f_{X}(x) &=& \begin{cases}
             0                                                  &  x < 0\\
             \sqrt{\frac{2}{\pi \sigma^2_x}} \exp{\left(-\frac{x^2}{2\sigma^2_x}\right)} & x \geq 0.\\
             \end{cases}
\end{eqnarray}
We see that $y_1$ and $y_2$ can be negative since both $z_1$ and $z_2$ in (\ref{noise_samples_in_rcv_signal_pos})
and (\ref{noise_samples_in_rcv_signal_neg}) have double sided Gaussian distributions.
The clipping process at the negative clipper is given by
\begin{equation}
      \label{clipped_signal_at_the_rcvr1}
       \tilde{y}_1 = [{y}_1]^+ \triangleq \begin{cases}
                          y_1 & \text{if } y_1\geq 0\\
                          0 & \text{otherwise}  \\
             \end{cases}
\end{equation}
\begin{equation}
      \label{clipped_signal_at_the_rcvr2}
       \tilde{y}_2 = [{y}_2]^- \triangleq \begin{cases}
                          y_2 & \text{if } y_2\geq 0\\
                          0 & \text{otherwise}.  \\
             \end{cases}
\end{equation}
Since $x$ is equal likely to appear in $y_1$ and $y_2$, for simplicity,
we assume $x$ appears in $y_1$ only. For a given $x$,
the equivalent noise power $\sigma^2_{\text{NC}}(x)$ can be computed as
\begin{eqnarray}
      \label{clipped_signal_at_the_rcvr}
         \sigma^2_{\text{NC}}(x)          &=& \frac{1}{2} E\left[\left(\tilde{y}_1 - \tilde{y}_2 - x\right)^2\right] \\
         \nonumber                        &=& \frac{1}{2} E\left[\left(\tilde{y}_1 - x\right)^2\right] + \frac{\sigma^2_z}{4} - \frac{\sigma_z}{\sqrt{2} \pi}E\left[\tilde{y}_1 - x\right]
\end{eqnarray}
where
\begin{eqnarray}
     \nonumber     E[(\tilde{y}_1 - x)^2] &=& \frac{\left(x^2-\sigma_z ^2\right)}{2} \text{erfc}\left(\frac{x}{\sqrt{2} \sigma_z }\right) -\frac{x \sigma_z e^{-\frac{x^2}{2 \sigma_z ^2}} }{\sqrt{2 \pi }}+\sigma_z ^2 \\
     \nonumber     E[\tilde{y}_1 - x]   &=& \frac{\sigma_z  e^{-\frac{x^2}{2 \sigma_z ^2}}}{\sqrt{2 \pi }}-\frac{x}{2} \text{erfc}\left(\frac{x}{\sqrt{2} \sigma_z }\right)
\end{eqnarray}
because $\tilde{y}_1$ and $\tilde{y}_2$ are independent.
Since $\sigma^2_{\text{NC}}(x)$ is conditioned by $x$,
we can estimate $\sigma^2_{\text{NC}}$ using
\begin{eqnarray}
    \label{eqivalent_noise_pwr_after_zc}
    \sigma^2_{\text{NC}}     &=& \int^{\infty}_{0} \sigma^2_{\text{NC}}(x) f_{X}(x) \, dx\ \\
    \nonumber                &=& \frac{{\sigma_z}^2}{2} + \frac{-\sigma_z \sqrt{{\sigma_z}^2+ {\sigma_x}^2} +\left( {\sigma_z}^2 + {\sigma_x}^2 \right) \text{tan}^{-1}\left(\frac{\sigma_z }{\sigma_x}\right)}{2 \pi}
\end{eqnarray}
Similarly, when $x$ contributes to $\tilde{y}_2$,
we simply obtain the same $\sigma^2_{\text{NC}}$ in (\ref{eqivalent_noise_pwr_after_zc}).
Hence, the total equivalent noise after negative clipping is
given in (\ref{eqivalent_noise_pwr_after_zc}) and the improved equivalent SNR per sample is
\begin{eqnarray}
    \label{SNR_improvement_eqn_zc}
        \text{SNR}_{\text{NC}} &=& \frac{E[x^2]}{\sigma^2_{\text{NC}}} ~=~ \frac{\sigma^2_x}{2 \sigma^2_{\text{NC}}}
\end{eqnarray}
Substituting $\text{SNR}_{\text{NC}}$ to (\ref{theroitical_BER_eqn}) yields the new theoretical BER performance
at the first stage of noise filtering.

\subsection{Threshold Based Noise Filtering Algorithm}
Without loss of generality, we assume that the signal component $x$ is only
contained in $y_1$, as in (\ref{clipped_signal_at_the_rcvr1}),
since the following development can be easily applied
to the dual case (i.e., $x$ is only contained in $y_2$, as in (\ref{clipped_signal_at_the_rcvr2})).

Under this assumption, after the first stage of noise filtering (negative clipper), we have
\begin{eqnarray}
 \label{Assump_threshol_algo}
   {\tilde{y}}_1 &=& [x + {z}_1(x)]^+ = x + {\tilde{z}}_1(x)\\
   {\tilde{y}}_2 &=& [{z}_2]^+ =\tilde{z}_2
\end{eqnarray}
where $\tilde{z}_1(x)$ and $\tilde{z}_2$ are the additive noise components after the negative clipper,
and the $[\cdot]^+$ operator is defined in (\ref{clipped_signal_at_the_rcvr1}).
Let $\tilde{y}$ be the reconstructed bipolar OFDM sample given by
\begin{eqnarray}
  \nonumber \tilde{y} & \triangleq & {\tilde{y}}_1 -  {\tilde{y}}_2 \\
                      &=& [x + {z}_1(x)]^+ - [{z}_2 ]^+ \\
                      &=& x + {\tilde{z}}_1(x) - {\tilde{z}}_2
\end{eqnarray}
We notice that $\tilde{z}_1(x)$ has a dependency on $x$ as a result of the negative clipping.
Since $x$ is always positive, there is a higher likelihood that the signal
sample ${\tilde{y}}_1$ containing $x$ is greater than the pure noise sample ${\tilde{y}}_2$.
Ideally, if we can perfectly identify the signal sample  ${\tilde{y}}_1$,
we can simply ignore the sample containing any noise, i.e., we set ${\tilde{y}}_2$ to $0$.
Hence, at the receiver, the binary decision is made by looking at the
difference between $\tilde{y}_1$ and $\tilde{y}_2$, relative to a threshold $c$.
All the possible outcomes are shown in Table \ref{y1_y2_cases}.
In Case A, the difference  $|{\tilde{y}}_1 - {\tilde{y}}_2|$ is below the threshold, so that
both $\tilde{z}_1(x)$ and $\tilde{z}_2$ contribute to the overall noise power of the output sample
${\tilde{y}} = {\tilde{y}}_1 - {\tilde{y}}_2$.
Case B corresponds to the ideal case discussed above and the output ${\tilde{y}} = {\tilde{y}}_1$
has a significantly reduced noise power.
Case C is the least frequent but will cause a completely wrong estimation of $x$.
In all cases an incorrect decision may not only destroy the signal, but also increase the noise power.
The corresponding algorithm is given below.

\begin{algorithm}
    \caption{: Threshold based noise reduction algorithm}
    \label{threshold_based_algo}
    \begin{algorithmic}
      \STATE {\bf Input} ${\tilde{y}}_1$, ${\tilde{y}}_2$, $c$
       \IF {${\tilde{y}}_1 - {\tilde{y}}_2 > c$}
                \STATE ${\tilde{y}}_2 \leftarrow 0$
       \ELSIF {${\tilde{y}}_2 - {\tilde{y}}_1 > c$}
                \STATE ${\tilde{y}}_1 \leftarrow 0$
       \ENDIF
       \STATE {\bf Output} ${\tilde{y}} ={\tilde{y}}_1 - {\tilde{y}}_2 $
    \end{algorithmic}
\end{algorithm}

We observe that the probability of each of the three cases
depends on $x$ and the threshold $c$.
In the next section, we will compute the optimal threshold to achieve the
best possible BER performance of Flip-OFDM.

    \begin{table}
    \renewcommand{\arraystretch}{1.4}
    \begin{center}
    \caption{Threshold based algorithm outputs}
    \begin{minipage}{\linewidth} \center
    \begin{tabular}{c|c|c}
    \hline
            Case  &                                                          &   Outputs ($\tilde{y}$)            \\
            \hline \hline
              A   & $| \tilde{y}_1 - \tilde{y}_2 |              \leq c $     &   $\tilde{y}_{1} - \tilde{y}_{2} = x + {\tilde{z}}_1(x) - {\tilde{z}}_2$  \\
              B   & $  \tilde{y}_1 - \tilde{y}_2                > c    $     &   $\tilde{y}_{1} = x + {\tilde{z}}_1(x)$        \\
              C   & $  \tilde{y}_2 - \tilde{y}_1                > c    $     &   $\tilde{y}_{2} = {\tilde{z}}_2$            \\
    \hline
    \end{tabular}
    \end{minipage}
    \label{y1_y2_cases}
    \end{center}
    \end{table}

    \begin{figure}[t!]
        \centering%
        \subfigure[$\tilde{y}_1$ and $\tilde{y}_2$ distribution if $x$ is fixed]{\includegraphics[scale=.55]{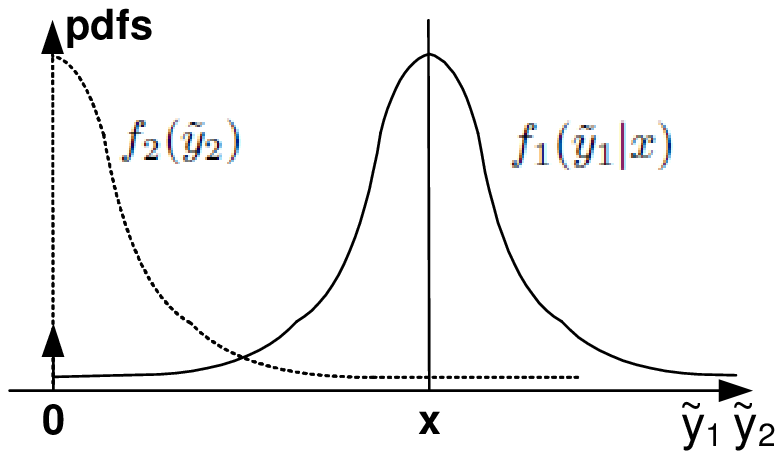}\label{y1andy2lbls}}
        \subfigure[Operation regions of the Algorithm]{\includegraphics[scale=.56]{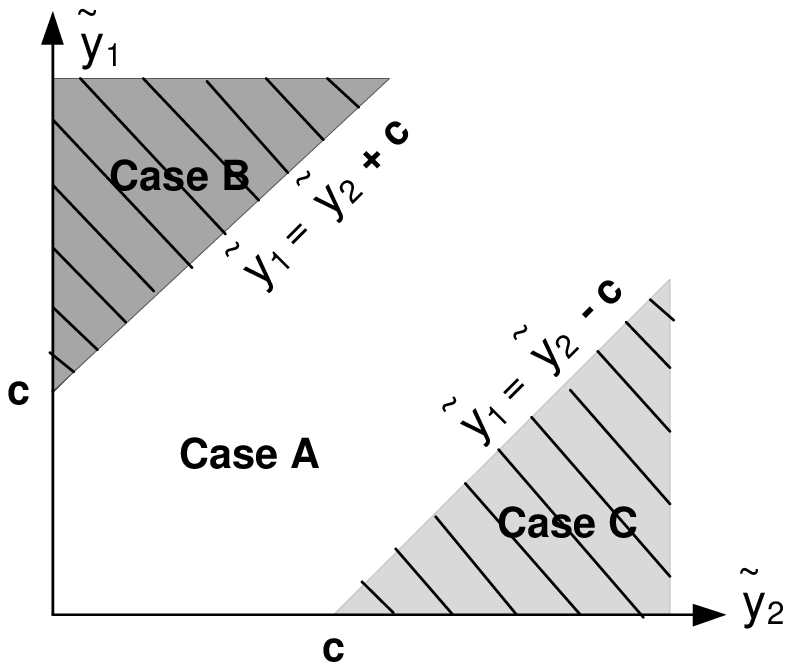}\label{algo_cases}}
        \caption{$\tilde{y}_1$ and $\tilde{y}_2$ distribution and the operation regions of the threshold based noise reduction algorithm}
        \label{y1andy2distribution}
    \end{figure}

\subsection{Design of the optimal threshold and performance analysis}
\begin{figure}[t!]
        \centering%
        \includegraphics[scale=.28]{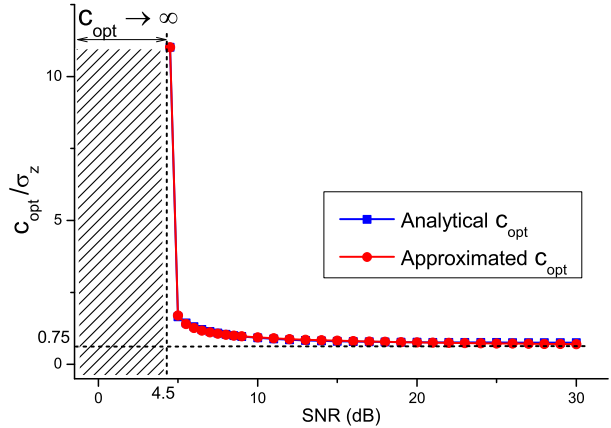}
        \caption{The variation of the theoretical optimum $c$ ($c_{\text{opt}}$) and approximated optimum $c$ ($\tilde{c}_{\text{opt}}$) with electrical SNR at the receiver. $c_{\text{opt}}$ and $\tilde{c}_{\text{opt}}$ are normalized with the standard deviation of the noise power}
        \label{lbl_optimum_c_vs_snr}
\end{figure}

The objective of the threshold base noise filtering algorithm is to minimize
the overall noise power by tuning the threshold $c$.
The algorithm differs from standard detection algorithms due to three
main reasons: (\emph{i}) $x$ is not fixed and has a single sided Gaussian distribution,
as given in (\ref{distribution_of_tx_signal}); (\emph{ii}) $\tilde{z}_1(x)$ and $\tilde{z}_2$
are not Gaussian random variables since they are clipped by the negative clipper;
and (\emph{iii}) $\tilde{z}_1(x)$ distribution depends on the random variable $x$.
In order to simplify the analysis, we first fix $x$, as shown in Fig. \ref{y1andy2distribution},
and find the equivalent noise power conditioned on $x$. That is, given $x$, the pdfs of $\tilde{y}_1$ and $\tilde{y}_2$ are
\begin{eqnarray}
    f_{1}({\tilde{y}}_1|x) &=&  \begin{cases}
             \frac{\delta({\tilde{y}}_1)}{2} \mbox{erfc}\left(\frac{x}{\sqrt{2} \sigma_z}\right) & \text{if } {\tilde{y}}_1 = 0\\
             \frac{1}{\sqrt{2 \pi} \sigma_z} \exp{\left(-\frac{({\tilde{y}}_1 - x)^2}{2\sigma^2_z}\right)} & \text{if } {\tilde{y}}_1 >0\\
             \end{cases} \\
    f_{2}({\tilde{y}}_2) &=&  \begin{cases}
             \frac{\delta({\tilde{y}}_2)}{2}                                                      & \text{if } {\tilde{y}}_2 = 0\\
             \frac{1}{\sqrt{2 \pi} \sigma_z} \exp{\left(-\frac{{\tilde{y}}_2^2}{2\sigma^2_z}\right)} & \text{if } {\tilde{y}}_2 > 0\\
             \end{cases}
\end{eqnarray}
where $f_{1}({\tilde{y}}_1|x)$ is a function of $x$.
Let $\sigma^2_{\text{A}}(c, \sigma_z, x)$, $\sigma^2_{\text{B}}(c, \sigma_z, x)$ and $\sigma^2_{\text{C}}(c, \sigma_z, x)$ be
the noise powers of Case A, B and C for a given $x$. Since $\tilde{y}_1$ and $\tilde{y}_2$ are independent, we obtain
\begin{eqnarray}
   \nonumber \sigma^2_{\text{A}}(c, \sigma_z, x) &=&  \int\limits_0^c \! \, \mathrm{d} \tilde{y}_1 \int\limits_0^{{\tilde{y}}_1+c} \! ({\tilde{y}}_1 - {\tilde{y}}_2 - x)^2 f_{1}({\tilde{y}}_1|x) f_{2}({\tilde{y}}_2) \, \mathrm{d} \tilde{y}_2 \\
   \nonumber                       &+& \int\limits_c^{\infty} \! \, \mathrm{d} \tilde{y}_1 \int\limits_{{\tilde{y}}_1-c}^{{\tilde{y}}_1+c} \! ({\tilde{y}}_1 - {\tilde{y}}_2 - x)^2 f_{1}({\tilde{y}}_1|x) f_{2}({\tilde{y}}_2) \, \mathrm{d} \tilde{y}_2 \\
   \nonumber \sigma^2_{\text{B}}(c, \sigma_z, x) &=& \int\limits_c^{\infty} \! \, \mathrm{d} \tilde{y}_1 \int\limits_0^{{\tilde{y}}_1-c} \! ({\tilde{y}}_1 - x)^2 f_{1}({\tilde{y}}_1|x) f_{2}({\tilde{y}}_2) \, \mathrm{d} \tilde{y}_2 \\
   \nonumber \sigma^2_{\text{C}}(c, \sigma_z, x) &=& \int\limits_0^{\infty} \! \, \mathrm{d} \tilde{y}_1 \int\limits_{{\tilde{y}}_1+c}^{\infty} \! ({\tilde{y}}_2 + x)^2 f_{1}({\tilde{y}}_1|x) f_{2}({\tilde{y}}_2) \, \mathrm{d} \tilde{y}_2 \\
\end{eqnarray}
as shown in Fig. \ref{y1andy2lbls}. The addition of the individual noise contribution in each case
corresponds to the average noise power of two time samples.
Hence, the average noise power per sample $\sigma^2_{\text{eq}}(c, \sigma_z, x)$ can be given as
\begin{eqnarray}
    \nonumber \sigma^2_{\text{eq}}(c, \sigma_z, x) &=&\frac{\sigma^2_{\text{A}}(c, \sigma_z, x) + \sigma^2_{\text{B}}(c, \sigma_z, x) + \sigma^2_{\text{C}}(c, \sigma_z, x)}{2}
\end{eqnarray}
and it is a function of $x$. Averaging over $x$, $\sigma^2_{\text{eq}}(c, \sigma_z, x)$ can be estimated as
\begin{eqnarray}
    \label{total_noise_eqn}
       \sigma^2_{\text{eq}}(c, \sigma_z, \sigma_x)  &=&  \int\limits_0^\infty \sigma^2_{\text{eq}}(c, \sigma_z, x) f_{X}(x) \,dx.
\end{eqnarray}
where $f_{X}(x)$ is given in (\ref{distribution_of_tx_signal}).
The optimum $c$, denoted by $c_{\text{opt}}$, can be selected such that
\begin{eqnarray}
     \label{opti_c_eqn}
        \frac{\partial \sigma^2_{\text{eq}}(c, \sigma_z, \sigma_x)}{\partial c} = 0.
\end{eqnarray}
Simply we can use numerical techniques to solve (\ref{opti_c_eqn}) to find $c_{\text{opt}}$.
Fig. \ref{lbl_optimum_c_vs_snr} shows the variation of the normalized $c_{\text{opt}}$ with $\text{SNR}$ in (\ref{exten_elec_snr_def}).
We see that, when $\text{SNR} <4.5$dB, $c_{\text{opt}}$ is infinite and the algorithm does not bring any gain. In such a case, we have
\begin{eqnarray}
     \lim_{c\to \infty} \sigma^2_{\text{eq}}(c, \sigma_z, \sigma_x) = \sigma^2_{\text{NC}}
\end{eqnarray}
where $\sigma^2_{\text{NC}}$ is given in (\ref{eqivalent_noise_pwr_after_zc}).
When $\text{SNR}> 4.5$dB, $c_{\text{opt}}$ is finite and there is a significant SNR gain.
Hence, the new SNR can be computed as
  \begin{eqnarray}
      \label{final_elec_snr}
         \text{SNR}_{\text{TOT}} & \simeq & \frac{\sigma^2_x}{\sigma^2_{\text{eq}}(c_{\text{opt}}, \sigma_z, \sigma_x)}.
  \end{eqnarray}


Since $c_{\text{opt}}$ only depends on $\text{SNR}$, we can approximate it as a function of $\text{SNR}$.
Using curve-fitting technique, we can approximate $c_{\text{opt}}$ using the following function
\begin{eqnarray}
    \label{approx_c_opt}
    \tilde{c}_{\text{opt}} / \sigma_z =  0.75  \frac{((\text{SNR} - 4.5)^n + a)}{((\text{SNR} - 4.5)^m +b)}
\end{eqnarray}
where $(a,b,m,n) = (0.9336, 0.03341, 0.4875, 0.3982)$.

\begin{figure}[t!]
        \centering%
        \includegraphics[scale=.30]{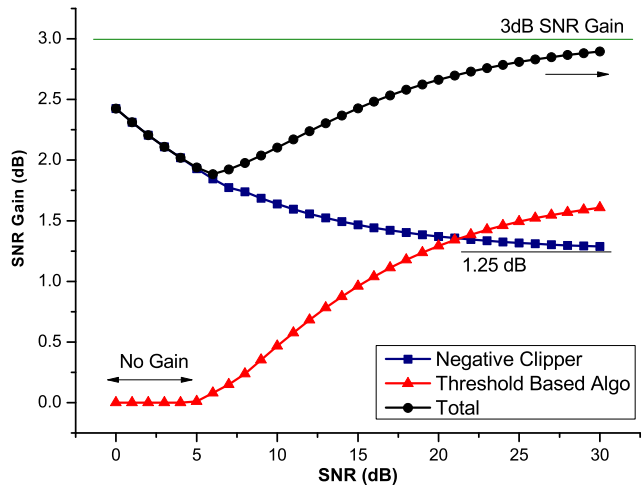}
        \caption{The SNR gains of the negative clipper and the threshold based noise filtering algorithm with electrical SNR at the receiver}
        \label{snr_gain_in_zc_algo}
\end{figure}

The SNR gains from (i) the negative clipper, (ii) the algorithm, and (iii) both negative clipper
and the algorithm are shown in Fig. \ref{snr_gain_in_zc_algo}, respectively.
The majority of the gain at low SNR ($< 4.5$ dB) is contributed by the negative clipper only.
However, the gain from negative clipper reduces as SNR increases; and stays at about $1.25$dB at higher SNR ($20-30$dB).
In contrast, the SNR gain from the algorithm increases steadily as SNR increases. At higher SNRs ($25-30$dB),
the overall SNR gain from both negative clipper and the algorithm is almost $3$dB.

\begin{figure}[t!]
        \centering%
        \includegraphics[scale=.60]{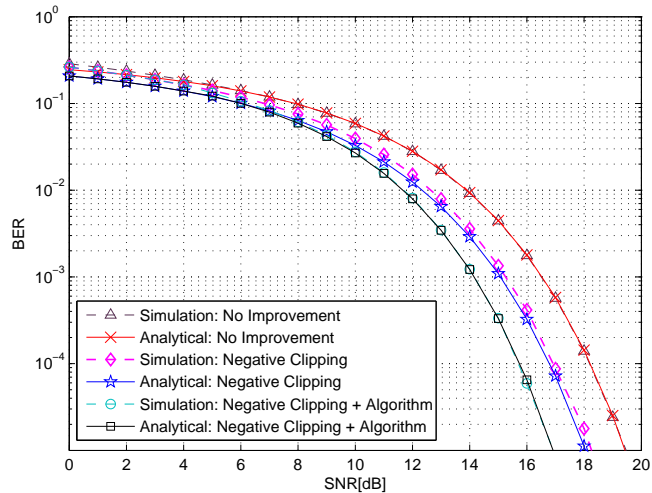}
        \caption{The improvements of BER performances (16-QAM) due to the negative clipper and the threshold based noise filtering algorithm.}
        \label{fig_ber_of_zc_thresh}
\end{figure}

In Fig. \ref{fig_ber_of_zc_thresh}, we compare both the simulated BER performance and the
theoretical one at each noise filtering stage, where $16$-QAM signalling is used at the transmitter.
Note that the noise after the negative clipper and the algorithm are no longer Gaussian.
However, for a large $N$, the FFT operation has the effect of whitening the noise in frequency domain.
Hence, the theoretical BER expression in (\ref{theroitical_BER_eqn}) is still valid,
given SNRs in (\ref{SNR_improvement_eqn_zc}) and (\ref{final_elec_snr}) for the respective noise filtering stages.

Observing Figs. \ref{snr_gain_in_zc_algo} and \ref{fig_ber_of_zc_thresh},
the SNR gains in Fig. \ref{snr_gain_in_zc_algo} are accurately reflected in the BER curves.
We see that, using the negative clipper and the proposed algorithm,
the system has $2.5$dB gain
at BER of $10^{-4}$, when compared to the original system.

\section{Conclusion}
We analyzed a unipolar OFDM technique (Flip-OFDM) for unipolar communication systems.
We showed that it is equivalent to the well-known ACO-OFDM in terms of spectral efficiency and error performance,
but can save nearly 50\% of receiver complexity over ACO-OFDM.
Moreover, we proposed a noise filtering algorithm used right after the negative clipper at the receiver.
Both negative clipper and the noise filtering algorithm can jointly contribute up to $3$dB gain at high SNRs.
Future work will focus on the potentials of Flip-OFDM for non-coherent RF wireless communications.


\bibliographystyle{IEEEtran}

\end{document}